# RAGVA: Engineering Retrieval Augmented Generation-based Virtual Assistants in Practice

Rui Yang[a,b], Michael Fu[c], Chakkrit Tantithamthavorn[a], Chetan Arora[a], Lisa Vandenhurk[b], Joey Chua[b]

[a]*Monash University, Australia*
[b]*Transurban, Australia*
[c]*The University of Melbourne, Australia*

**Abstract**

Retrieval-augmented generation (RAG)-based applications are gaining prominence due to their ability to leverage large language models (LLMs). These systems excel at combining retrieval mechanisms with generative capabilities, resulting in contextually relevant responses that enhance user experience. In particular, Transurban, a road operation company, replaced its rule-based virtual assistant (VA) with a RAG-based VA (RAGVA) to offer flexible customer interactions and support a wider range of scenarios. This paper presents an experience report from Transurban's engineering team on building and deploying a RAGVA, offering a step-by-step guide for creating a conversational application and engineering a RAGVA. The report serves as a reference for future researchers and practitioners. While the engineering processes for traditional software applications are well-established, the development and evaluation of RAG-based applications are still in their early stages, with numerous emerging challenges remaining uncharted. To address this gap, we conduct a focus group study with Transurban practitioners regarding developing and evaluating their RAGVA. We identified eight challenges encountered by the engineering team and proposed eight future directions that should be explored to advance the development of RAG-based applications. This study contributes to the foundational understanding of a RAG-based conversational application and the emerging AI software engineering challenges it presents.

## 1. Introduction

In the rapidly evolving modern digital landscape, virtual assistants (VAs) have emerged as pivotal tools for enhancing user interactions and operational efficiency [1]. From streamlining customer service operations in e-commerce to optimising workflow management in enterprise environments, VAs are revolutionising the way digital companies engage with their clients and manage internal processes. Most of the current VAs are *rule-based* (or task-oriented), commonly known as chatbots, and operate based on predefined rules to handle specific scenarios [2]. Traditional VAs mostly use a decision tree menu to address common scenarios to support users or answer their queries. Despite their widespread adoption, traditional VAs often fall short in handling queries that deviate (even) slightly from predefined patterns [3], maintaining contextual accuracy, and often restrict interactions to menu selections, posing significant challenges for IT professionals and software engineers striving to develop more advanced and reliable systems.

The latest trend in the software industry is the integration of artificial intelligence (AI) capabilities [4, 5, 6]. The development of AI has been significantly advanced by Large Language Models (LLMs) [7] such as GPT-4 [8], Claude [9], and LLaMa [10], which have substantially improved the machine's ability to understand and generate human-like language. In particular, retrieval-augmented generation (RAG)-based virtual assistants are transforming automated user support interactions [11]. RAG combines LLMs' generative capabilities with information retrieval, allowing LLMs to access and integrate external knowledge sources, e.g., domain-specific documents, and alleviate hallucination issues [12]. RAG-based applications are gaining traction in several software engineering tasks [13, 14, 15, 16]. For VAs, RAG offers means to enhance the accuracy and contextual relevance of customer support responses, effectively addressing the limitations of rule-based assistants, and hence RAG-based solutions are being deployed to build VAs in different application domains, such as, medicine [17, 18] and education [19].

Engineering a RAG-based application, such as a virtual assistant, presents emerging and significant software engineering challenges that demand the attention of SE practitioners [20, 21]. Developing and evaluating these applications is inherently complex due to the integration of nondeterministic LLMs, which exhibit unpredictable behaviors. Unlike traditional software applications, where behavior is more controllable and requirements can be clearly defined, RAG-based applications necessitate continuous validation and adaptive development strategies. Furthermore, the principles of Responsible AI (RAI) introduce additional complexity in evaluating RAG-based systems, particularly in the context of LLMs. Handling key aspects such as hallucinations, faithfulness, bias, toxicity, contextual precision/recall, contextual relevancy, and knowl-edge retention requires rigorous assessment [22, 23, 24, 25]. Nevertheless, these emerging challenges in engineering a RAG-based virtual assistant remain largely uncharted in the literature, hindering clear pathways for developing reliable, effective, and ethically sound RAG-based applications.

In this experience report, we explore the challenges and the future directions of developing RAGVAs in practice, based on our experiences with our industry partner - *Transurban*, which is an Australia-based company that builds and operates toll roads in Australia, the United States and Canada and is the world's largest toll operator. Transurban provides several services, such as electronic-TAG (e-TAG), for toll road travel payments and implements several software solutions, e.g., VAs for toll road-related customer service. In collaboration with a group of practitioners at Transurban, we offer a detailed step-by-step guide, challenges and future directions for constructing a conversational application and engineering a RAG-based system, drawing from their practical experiences. This comprehensive guide provides insights and a roadmap for developing a RAG-based conversational application that serves as a valuable resource for future researchers and practitioners. We report on the outcomes of a multi-day focus group conducted with *nine* Transurban engineering team members to identify and analyze key software engineering challenges in developing and evaluating their RAGVA. These insights highlight critical areas for future research and provide a foundation for advancing this critical and burgeoning field.

***Contribution.*** Based on our focus group findings with the Transurban engineering team, the contributions of this paper are as follows:

- A comprehensive guideline for developing a Customer Virtual Assistant.

- A comprehensive engineering framework for building a Retrieval-Augmented Generation-based Virtual Assistant (RAGVA), covering all stages from implementation to testing.

- Identification of eight key challenges in engineering RAG-based software applications, accompanied by 22 research questions to outline concrete directions for future work.

***Structure.*** The remainder of the paper is structured as follows. Section 2.1 provides a background on the rule-based VAs and their limitations. Section 3 delineates RAGVAs at Transurban. Section 4 provides details of our focus group settings. Section 5 details the outcomes of our focus group in the form of challenges and future research directions while developing RAGVAs, Section 8 discloses limitations, and Section 9 concludes the paper.

## 2. Background and Related Works

### 2.1. Rule-Based Virtual Assistant

Linkt is Transurban's e-TAG tolling brand in Australia [26]. Linkt customers often seek assistance through its website for aspects such as toll road information, payments and toll invoices, account and billing issues, and avoiding scams and fraud. In this section, we begin by discussing the motivation for implementing a VA to support the customer service team at Linkt. We then provide a step-by-step guide for businesses looking to develop and deploy a customer VA, drawing on the experience of Transurban's engineering team. This guide applies to both rule-based and AI-based VAs, offering a general workflow and discussing critical aspects to consider when building a VA for customer service. Finally, we present the limitations of the existing rule-based VA at Transurban and discuss how large language models (LLMs) can address them.

#### 2.1.1. Motivation

Transurban developed a VA for Linkt to enhance customer experience and operational efficiency. This solution provides immediate assistance for toll road information, payments, invoices, and account management, reducing wait times and ensuring consistent service. Customers often spend valuable time navigating complex content on Linkt's website to pay tolls. The VA streamlines this process by swiftly guiding them to the payment page, enhancing convenience and satisfaction. Ultimately, VA aims to deliver more personalized and efficient tolling services that cater directly to customer needs.

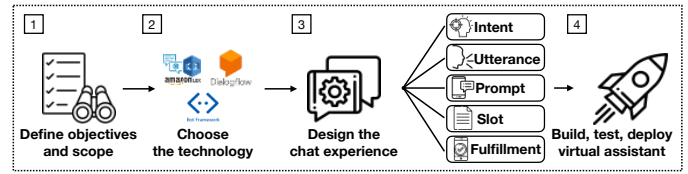

Figure 1: The four-step process for developing a customer VA.

#### 2.1.2. Developing a Customer Virtual Assistant: A Step-by-Step Guide

Figure 1 outlines the four-step process (elaborated below) of developing a customer VA.

**Step 1 : Define Objectives and Scope.** To create a VA that effectively addresses customer needs, it begins by conducting comprehensive user research. This involves analyzing data collected from various sources, including their customer service centre, on-site service interactions, and call centre logs. By examining common inquiries and customer issues, the most frequent and pressing concerns can be identified. Through this data-driven approach, Transurban has identified the main use cases as travel payment, travel expense estimation, customer e-TAG management, toll invoice management, and customer account management. This allows them to pinpoint specific pain points and recurring questions, ensuring the VA addresses the key areas of customer need.

**Step 2 : Select the VA Platform.** There are several platforms, such as Amazon Lex [27], Google Dialogflow [28], and Microsoft Bot Framework [29], which facilitate enterprise development and deployment of customized VAs. When the engineering team evaluated different available platforms, they considered several critical factors to ensure the chosen solution met their specific needs. Integration capabilities were a top priority, as the assistant needed to seamlessly integrate with Transurban's existing systems, including their website, mobile app, and customer relationship management (CRM) software. Ease of use was also crucial; the team sought a platform that offered an intuitive interface and robust support documentation to facilitate swift development and deployment. Additionally, they assessed the platforms' ability to handle high volumes of concurrent interactions without performance degradation.

**Step ③: Design the Chat Experience.** The development of the static rule-based customer VA began with the engineering team designing a main menu that outlines each identified use case from Step ①. This is followed by crafting user journey maps for each use case, delineating the anticipated interactions between users and the assistant. This iterative process converged into a rule-based VA structured like a decision tree-like menu system. Upon initiation, users are greeted with a welcoming message and presented with a main menu interface. Each menu option directs the user through subsequent sub-menus until task completion.

The engineering process of designing user journey maps is multifaceted, requiring careful consideration of five key elements: intent, utterance, prompt, slot, and fulfillment. Below, we introduce each element through a specific use case at Transurban, where a customer intends to pay for their recent travel using the VA:

*Intent* refers to the action the user wants to accomplish. In our example, the intent is for the customer to "*Pay for my travel*." Recognizing this intent is crucial, as it drives the conversation flow and determines which options are presented in the main menu.

*Utterance* refers to the actual input from the user in the form of selections from the predefined options in this case. When the customer chooses "*Pay for my travel*," it initiates the relevant pathway in the assistant. Designing effective menu options requires anticipating user intent and presenting them clearly.

*Prompt* is a message or question from the VA that encourages the user to provide more information or clarify the request. In Transurban's context, it meant guiding the user through menu selections to gather necessary information. After recognizing the intent, the assistant offers options such as "*Pay for a recent trip*," "*Ongoing travel*," or "*Single trip*." Clear prompts ensure users can easily navigate the system and understand the choices available.

*Slot* is a placeholder for the information the assistant needs to fulfil the intent. For example, once the user selects "*Pay for a recent trip*," the slot could be the specific trip details required to process the payment. Designing slots involves determining necessary information and how to collect it efficiently.

*Fulfillment* is the completion of the user's desired action. After selecting the appropriate payment option, the assistant provides a link for online payment and a final fulfillment message offering options like "*Main menu*," "*Help center*," or "*Chat to someone*." This ensures that the interaction is concluded satisfactorily while offering further assistance if needed.

**Step ④: Build, Test, and Deploy the Virtual Assistant.** After designing the chat experience, the engineering team implemented the customer support menus within the system. Before deployment, they conducted thorough testing to ensure functionality and user satisfaction. They began with unit tests for each menu option and corresponding dialog, verifying that all pathways operated as intended. Subsequently, they engaged a group of customers for user testing, allowing real-world feedback to guide refinements. This continuous improvement phase helped identify potential issues and enhanced the VA's overall effectiveness, ensuring it met user expectations and business objectives. Once testing was complete, the team deployed the VA across the intended channels, such as their website and mobile app.

*2.1.3. Limitations of Existing Rule-Based Virtual Assistants and the Role of LLMs*

In the section above, we have presented a step-by-step guide for developing a customer service-focused VA, drawing on insights from Transurban's engineering team. Despite automating numerous toll road-related functions at Linkt, the current rule-based system exhibits notable limitations. Specifically, the Transurban team pinpointed three primary limitations encountered with the rule-based assistant during its tenure. In the subsequent sections, we illustrate each limitation within the context of Transurban and describe the role of large language models (LLMs) in mitigating these challenges.

**Limitation ①: Difficulty in Finding Relevant Information.** Unlike utility or telecommunication billing services with regular customer transactions, tolling transactions occur infrequently, e.g., once a year. This is because customers do not frequently change their vehicles. The infrequent nature of tolling transactions means customers often need more support due to their unfamiliarity with the system. At Linkt, around 96% of interactions are completed through digital channels. Automating responses for common queries can reduce the load on human agents. Thus, Linkt built a designated help page and a traditional VA using a decision tree menu to achieve this. However, customers sometimes still struggled to find relevant information or spent a lot of time, trying to interact with this rule-based VA.

**Limitation ②: Inconsistencies in User Experience and Limited Human Assistant Availability.** Traditional rule-based VAs at Linkt faced critical challenges in achieving seamless integration across web and app platforms while ensuring synchronized user experiences. These systems typically operate on rigid decision trees and predefined rules. Synchronizing interactions between web and app environments requires extensive manual intervention, leading to potential inconsistencies in user experience. Furthermore, the limited availability of human assistant support at Linkt further underscores the need for a robust automated solution to enhance customer engagement and satisfaction.

**Limitation ③: Insufficient Assistant Scalability and Adaptability.** As noted previously, the rule-based VA operates on a task-oriented framework, which poses challenges in scaling and addressing a broader spectrum of user needs as they evolve. For example, the integration of a new toll payment method necessitates extensive manual updates to the decision tree of the traditional VA to accommodate novel queries and processes. Maintaining and updating the decision tree menu in such a VA is also labor-intensive. Furthermore, the capacity of the VA to handle inquiries is constrained by the finite space within its menu structure, aiming to avoid overwhelming customers with an excessive array of choices.

*2.2. Related Works*

Recently, deep learning-based VAs have gained traction as the most advanced types of intelligent chat agents. Skrebeca et al. [30] classified such self-learning agents as either retrieval-based, a neural network to assign scores and select the most likely response from a set of responses, or generative-based, which synthesizes the response using deep learning techniques. Attigeri et al. [31] compared the performance of technical university information chatbots built using different types of NLP models, and found that neural network-related models outperformed other models such as TD-IDF and pattern-matching models. Vu et al. [32] presented an FAQ chatbot for online customer support that integrates retrieval-based techniques, such as a FAQ agent and Google Custom Search, with generative capabilities like question generation for augmenting training data. This hybrid approach leverages retrieval models to ground responses in factual data while utilizing generative models to address gaps in the knowledge base and enhance dialogue versatility.

Compared to the hybrid-based approach, RAG-based systems offer more advanced capabilities through the use of LLMs. This hybrid architecture addresses key limitations of purely generative models, such as hallucination and factual inaccuracies, by anchoring responses in external knowledge. These RAG-powered chatbots have been utilized in a wide range of use cases, from educational to commercial to internal resources [33, 34, 35].

Existing research on RAG-based virtual assistants (VAs) has primarily focused on technical advancements, such as retrieval algorithms and generative response mechanisms [36, 37, 38]. While empirical evidence establishes RAG-based VAs as the state-of-the-art in question-answering benchmarks [11, 39], limited research provides actionable guidance for practitioners seeking to build such systems. To address this gap, our paper offers a novel software engineering perspective on the development of RAG-based VAs for customer service, drawing from the practical experiences of our industry partner—a road operation company. Specifically, we present a step-by-step guide for constructing a RAG-based VA and identify eight key software engineering challenges encountered during its development along with recommended future research directions.

## 3. RAGVA: Retrieval Augmented Generation-Based Virtual Assistant

In customer VA, RAG can efficiently retrieve information about toll road conditions and payment options, providing users with accurate, contextually relevant responses, surpassing the capabilities of a rule-based assistant. This integration enables RAG to generate more specific, diverse, and factual language by dynamically retrieving pertinent information from external sources (e.g., customer support documents).

In the context of a RAGVA for toll road customer service, this approach effectively addresses the three major limitations found in the existing rule-based virtual customer assistant. To address Limitation ①, RAG employs a vector store and prompt augmentation, enabling the assistant to retrieve pertinent information from a customer support dataset, thus helping customers find relevant information more quickly and accurately. For Limitation ②, the LLM in RAG can handle a broader range of scenarios compared to the rule-based assistant, which is limited to predefined scenarios, thereby reducing reliance on human assistance and enhancing user experience. Regarding Limitation ③, the flexibility of LLM in RAG allows the VA to scale more effectively and adapt to new situations, accommodating growing customer service demands and evolving requirements.

We present the RAGVA in Figure 2. This framework encompasses three major steps, from data preparation to evaluation. In the following sections, we first introduce Step ①, the data ingestion workflow, which prepares a store used for information retrieval to enhance user prompts and improve LLM responses. Next, we detail Step ②, the RAG-based response generation workflow, which outlines the process from customer input through information retrieval to response generation using the augmented prompt. Finally, we describe Step ③, which focuses on the evaluation workflow for assessing the RAG system.

As the aim of this section is to present a broader overview of developing RAGVA, rather than explaining the specifics of the methodologies from Transurban, we do not disclose the detailed development process that Transurban's RAGVA underwent due to confidentiality reasons.

*3.1. Data Ingestion Workflow*

The data ingestion workflow is summarized in Step ① of Figure 2. Transurban's engineering team began by collecting toll-related documents from the Linkt support website. One such document is titled *How to Pay Your Toll Invoice*, which contains multiple sections like payment methods, deadlines, and FAQs.

Given the variety of customer support scenarios, the raw documents presented challenges in their unstructured form. Thus in Step ①b, the engineering team conducted several data engineering activities to structure the data effectively. For instance, the FAQ section was identified as a critical part for retrieval. Metadata tags were added to the extracted content such as Content: Billing Support, Type: FAQ and Keywords: Toll Payment, Invoice, Deadlines. This tagging ensures the data is easy to categorize and retrieve later.

In Step ①c, the structured data were segmented into manageable document chunks. Instead of arbitrary segmentation, the document was divided based on its structure, as shown by the chunks in Figure 3. Each chunk represents a self-contained and retrievable unit. Given that some documents may contain image elements, the system uses an image-to-text model to con-vert these images into text before chunking, as shown in Figure 3. We further discuss how to select the appropriate chunking strategies in Section 7.1.

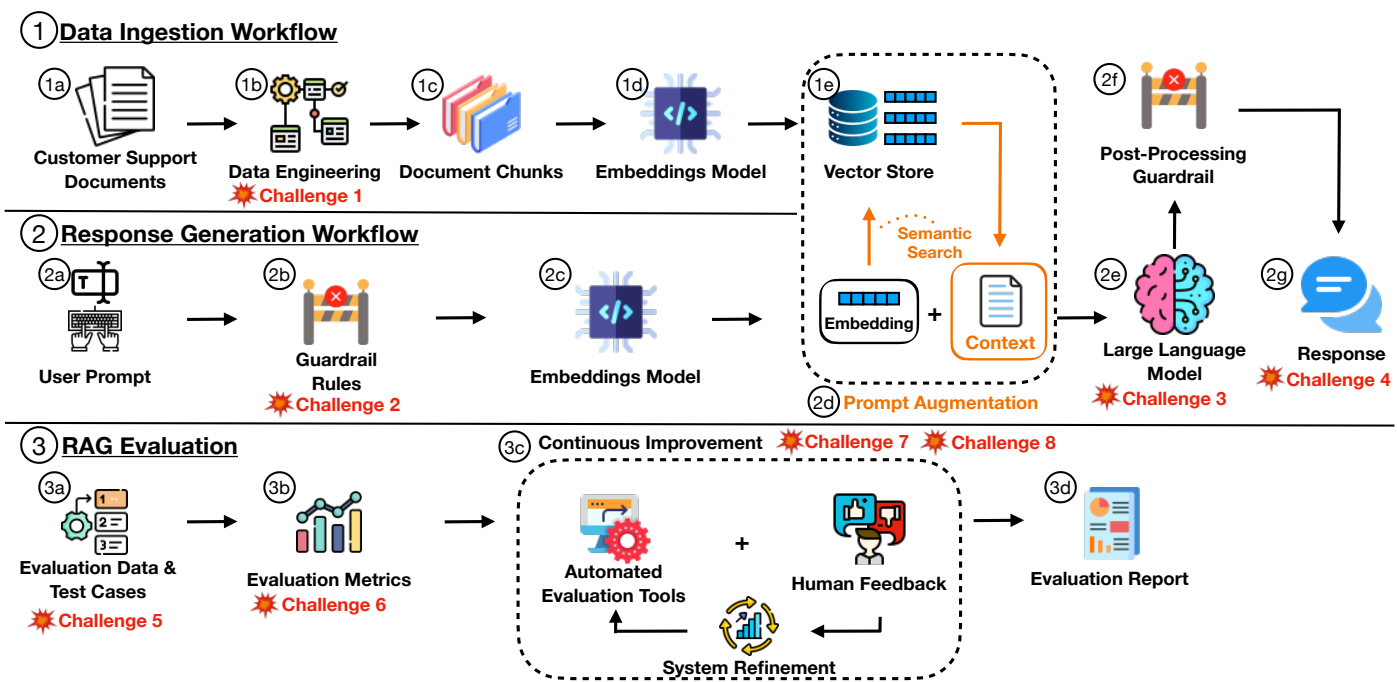

Figure 2: An overview of the Retrieval-Augmented Generation (RAG)-based VA developed for customer support. Step 1 details the data preparation process for building a vector knowledge base used for prompt augmentation. Step 2 details the response generation workflow using prompt augmentation with a large language model (LLM). Step 3 details the automated evaluation process for assessing the RAGVA.

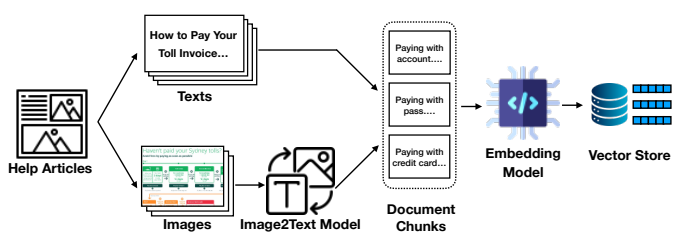

Figure 3: Example of multi-modal documents passing through the data ingestion pipeline.

In Step (1d), the document chunks with metadata tags were then processed by an embedding model, which transformed the text data into numerical vectors. These vectors were stored in a vector database in Step (1e), enabling efficient retrieval and analysis through indexing and matching user queries with the most appropriate semantically similar documents in the dataset. The selection of embedding models and vectoring databases is further discussed in Section 7.2.

Through these steps, the engineering team ensured that the customer support data was ready and accessible to support the RAG workflow.

### 3.2. RAG-Based Response Generation

The response generation workflow is summarized in Step ② of Figure 2.

Step ② of Figure 2 illustrates how this ingested data is retrieved and used to generate customer-centric responses. Consider a customer who asks, "How can I pay my toll invoice?" In Step (2b), this input is immediately passed through a set of guardrail rules designed to filter and sanitize the prompt, ensuring that inappropriate or irrelevant inputs are flagged or modified before processing further. Sensitive information such as personal details and billing details are also masked and anonymized in this step to protect user privacy. These steps are crucial to maintaining the integrity and quality of the system's responses, as well as complying with important data protection laws such as the GDPR [40].

Once the user prompt is sanitized, an embedding model is employed in Step (2c) to transform the user input into dense vector representations which encapsulates the semantic meaning of the input from Step (2a). In Step (2d), the embeddings are utilized in the prompt augmentation process, where additional context is retrieved from the vector store established in Step ①. In this case, the system identifies the "Payment Methods" and "FAQs" chunks as the most relevant results. These retrieved chunks are then combined with the user query to form an enriched prompt. For instance, the system constructs the following input for the large language model (LLM): "User asked: How can I pay my toll invoice? Related content: Payment Methods include Credit/Debit Card and PayPal."

In Step (2e), the enriched input is introduced into the LLM, which generates a response based on the augmented customer prompt in Step (2g). The LLM utilizes its extensive pre-trained knowledge and generates a response that might read, "You can pay your toll invoice using a credit or debit card. Please visit [Linkt Payment Page]. For more details, refer to our FAQs." As an additional step to remove any unwanted content in the responses generated by LLMs, post-generation guardrails can be deployed as shown in step ②f. State-of-the-art runtime solutions include Meta's LlamaGuard [41] and OpenAI's Moderation API [42], which utilize task-specific, pre-trained LLMs to classify responses as 'safe' or 'unsafe' based on the input or output. Nvidia's NeMo Guardrails [43] offers a customizable framework where developers can use rail files to run multiple guardrails simultaneously, addressing various harm categories and enabling engineers to define context-specific rules for evaluating response safety.

This RAG framework effectively addresses the knowledge-intensive nature of toll road customer support, where numerous support scenarios and an extensive range of possible responses exist. By leveraging ingested customer support documents with a powerful LLM, the system helps the model better understand diverse queries and guides customers more accurately to the correct solutions.

*3.3. RAG Evaluation*

Table 1 provides a comprehensive summary of the three critical evaluation aspects for assessing a RAG-based application, such as VA for toll road customer service.

Firstly, general benchmarking evaluates the components described in Steps ②d and ②e in Figure 2, focusing on assessing the generative capability and retrieval accuracy of a RAG system. This step ensures that the employed LLMs can manage general human question-answering scenarios and maintain a robust embedding and retrieval system.

Secondly, last-mile testing evaluates domain-specific retrieval and generation capabilities. For example, the test cases may include assessing the application's ability to assist customers with issues like toll payment and vehicle management. This step is labor-intensive, requiring significant engineering effort and collaboration between the customer support and engineering teams to develop relevant and useful test cases for diverse customer support scenarios. Additionally, this step includes a critical Responsible AI (RAI) evaluation, leveraging automated tools and metrics to ensure the system does not generate irrelevant, toxic, or biased outputs, thus promoting the responsible and ethical use of AI.

Lastly, application testing focuses on real-world scenarios after deployment, incorporating customer feedback and system performance assessment. This includes evaluating if the application meets customer needs and conducting tests to ensure smooth and stable system operation.

The RAG evaluation workflow is summarized in Step ③ of Figure 2. In Table 2, we illustrate each step in the RAG evaluation workflow with a representative example testing scenario: *"a customer who has recently traveled on a toll road and wishes to make their payment using the RAGVA."* In this testing scenario, the VA should provide clear instructions on how to pay for the toll, including necessary details such as payment methods, deadlines, and potential fees. Additionally, it should guide the customer to a specific payment page for secure transactions, where they can use credit or debit cards. This scenario is both common and crucial as it directly pertains to the primary service offered by Transurban and significantly impacts customer satisfaction.

*3.3.1. Evaluation Data & Test Cases*

In Step ③a of Figure 2, the engineering team focuses on gathering and organizing the data needed to simulate realistic customer interactions with their RAGVA. Considering our example testing scenario, the team begins by manually creating a comprehensive set of test cases for this scenario. These test cases include various prompts with similar meanings to test critical aspects of the VA introduced in Step ③ in Figure 1, such as intent recognition, utterance interpretation, prompt handling, slot filling, and fulfillment of the payment process. We provide example test cases for test cases for evaluating these aspects in Table 2, and further discuss test case generation methods in Section 7.3. By crafting detailed descriptions of each interaction and defining the expected outcomes, the team ensures that the assistant is evaluated against realistic and meaningful scenarios.

*3.3.2. Evaluation Metrics*

Step ③b of Figure 2, preparing evaluation metrics, is a critical step that ensures the system performs effectively and responsibly. This step focuses on both general benchmarking and last-mile testing. By employing a combination of performance-related and RAI metrics as summarized in Table 1, the evaluation team can comprehensively assess the assistant's capabilities and ensure it meets the high standards required for customer service scenarios.

Following our example testing scenario where a customer seeks to pay for recent toll road travel, generative capability metrics such as BLEU [44], ROUGE [45], and METEOR [46] assess the assistant's ability to accurately and fluently generate responses that guide customers on payment methods and instructions. In particular, BLEU evaluates the linguistic accuracy of the assistant's responses, ensuring they align closely with expected payment-related queries. ROUGE assesses how well the chatbot summarizes payment instructions, while METEOR ensures the responses are both coherent and comprehensive, helping maintain clarity in customer interactions.

In addition to performance-related metrics, RAI metrics are essential for ensuring the assistant's responses are reliable and ethical. Metrics shown in Table 1 help mitigate risks in customer interactions. For example, these metrics ensure the assistant provides accurate payment-related information without generating false or misleading details (hallucinations), avoids biased responses, maintains respectful language (toxicity), and retains relevant context throughout the conversation. This comprehensive evaluation approach ensures that the assistant not only performs effectively but also upholds ethical standards in customer service interactions.

*3.3.3. Continuous Improvement*

Continuous improvement is a cornerstone in refining the RAGVA. This ongoing process begins with the use of automated evaluation tools (such as those presented in Table 1) that rigorously test the assistant's responses across a variety of real-world scenarios. These tools simulate customer interactions, such as inquiries about payment processing or account status, to uncover inaccuracies, slow response times, and unhelpful outputs. The automated evaluation pinpoints specific areas where the system falls short. This phase ensures that the assistant is consistently evaluated against dynamic and practical test cases, maintaining high relevance to actual user needs and behaviors.

Table 1: An overview of retrieval-augmented generation (RAG) evaluation including general benchmarking, last-mile testing, and real-world application testing.

| Aspect | General Benchmarking | Last-Mile Testing | Application Testing |
|---|---|---|---|
| **Evaluation Focus and Objective** | Evaluating basic **generative capabilities** and **retrieval accuracy**; Achieving high performance on standard datasets | Ensuring domain-specific retrieval and response generation; Prioritizing **Responsible AI (RAI)** including adaptability to domain-specific queries and robustness | Testing virtual assistant performance in real-world customer service scenarios; Validating performance in operational settings, focusing on **customer-centric evaluation** and **system performance** |
| **Evaluation Data and Test Cases** | Static, pre-defined datasets | Human-crafted domain-specific test cases | Real-world, domain-specific interactions, live user data |
| **Evaluation Metrics** | **Generative Capability:**<br>BLEU [44]<br>ROUGE [45]<br>METEOR [46]<br>Annotation-Based Precision and Recall [47]<br><br>**Retrieval Accuracy:**<br>nDCG@k [48]<br>(Normalised Cumulative Discount Gain) | **RAI:**<br>Hallucinations [49]<br>Faithfulness [50]<br>Bias [51]<br>Toxicity [52]<br>Contextual Precision/Recall [53, 54]<br>Contextual Relevancy [55]<br>Knowledge Retention [56] | **Customer-Centric:**<br>Customer satisfaction<br>Resolution rate<br>Escalation Rate<br>User Engagement<br><br>**System Performance:**<br>Response time |
| **Real-World Relevance** | Moderate | High, closer to practical use cases | Very high, directly relevant to deployment contexts |
| **Examples** | **Generative Capability Benchmark:**<br>Natural Questions [47]<br>SQuAD 1.1 [57]<br>SQuAD 2.0 [58]<br><br>**Retrieval Benchmark:**<br>BEIR [59] | **Automated Evaluation Tool Support:**<br>DeepEval [60]<br>RAGAS [61]<br>LastMileAI [62] | **Customer-Centric:**<br>Customer Satisfaction Surveys<br>Real-time Customer Interaction Logs<br>Live User Feedback<br><br>**System Performance:**<br>System Uptime Reports |

Table 2: Test cases for the example testing scenario, "a customer wishes to make the payment using Transurban's RAG-based virtual assistant", to evaluate the five critical aspects of the virtual assistant.

| Test Aspect | Example Test Case | Expected Response/Outcome |
|---|---|---|
| **Intent Recognition** | Prompts like "How do I pay for my recent toll?", "I need to settle my toll charges", and "What's the process to pay my toll fee?" | The assistant correctly identifies the customer's intention and responds with: "To pay for your toll, please provide your vehicle's license plate number and the toll road you used. I will then guide you through the next steps." |
| **Utterance Interpretation** | Variations prompts like "I want to pay for the toll road", "I traveled on a toll road and need to pay", and "Can you help me pay my toll fee?" | The assistant accurately processes these variations without changing the response, maintaining consistency in asking for necessary details. |
| **Prompt Handling** | Customer inputs information for paying their toll, "License plate: ABC-123; Toll road: Melbourne M3 Highway" | The assistant interprets the customer's input correctly and responds with: "May I confirm that you wish to pay for the vehicle with license plate ABC-123, which recently traveled on the M3 highway in Melbourne?" |
| **Slot Filling** | Customer inputs information for paying their toll, "License plate: ABC-123; Toll road: Melbourne M3 Highway" | The assistant correctly captures and stores the required details for searching the system. |
| **Fulfillment** | Customer confirms "Yes, that is correct. I want to pay now" | The assistant confirms the trip and balance with the customer and then provides a link to the payment system with pre-filled payment information based on the details provided by the customer. The assistant guides the customer to the payment link and ensures they understand how to complete the transaction securely. |

On the other hand, user feedback can be collected from live customers after deployment and from focus groups before deployment. This feedback focuses on aspects such as accuracy, relevance, and clarity. Unlike automated tools, human feedback provides insights into user satisfaction. Furthermore, the assistant's responses are judged by human agents instead of LLM agents used in automated evaluation tools like RAGAS [61]. This helps pinpoint issues like misunderstanding human needs and context. The evaluation results from automated tools, combined with collected human feedback, will be analyzed to identify patterns and root causes of any deficiencies. Engineers perform root cause analysis, prompt refinement, update guardrail rules, and tune hyperparameters. This continuous refinement enhances the assistant's ability to handle domain-specific queries with greater accuracy and reliability.

## 4. Focus Group Settings

Given that RAG-based systems are still emerging within software engineering (SE), numerous unresolved engineering challenges impede their adoption in enterprises [21, 20]. This paper aims to explicate these issues through an SE lens by examining the SE challenges encountered by Transurban in developing their RAGVA for Linkt's toll road customer service. By drawing on Transurban's experience, we aim to identify key SE challenges in engineering RAG-based systems, review relevant literature, and recommend the direction of future research to address these challenges. In particular, we collected data through a focus group [63] with nine practitioners from the Transurban engineering team, conducted over two sessions. Each session lasted between 3-3.5 hours and focused on understanding the RAGVA development practices at Transurban and the challenges and directions for the engineering team. The focus group also included the research team (the first four co-authors). Our study design also obtained ethical approval from the Monash University Human Research Ethics Committee (MUHREC ID: 43572).

Because the practice of engineering and integrating RAG-based systems, such as a virtual assistant at Transurban, is still emerging and not uniform across all teams, there is no systematic way to identify the key stakeholders for the adoption of such technologies. To identify key stakeholders, we employed a combination of convenience and snowball sampling strategies. We used our immediate contacts at Transurban (the two co-authors from Transurban), who helped us recruit further relevant participants for the focus group. This approach allowed us to leverage initial contacts to refer additional relevant participants, ensuring a comprehensive understanding of the SE challenges and procedures involved in developing a RAG-based system at Transurban.

We conducted a focus group with nine members of Transurban's engineering team to gain a comprehensive understanding of the challenges and future directions in developing RAG-based virtual assistants. The focus group was designed to facilitate an open and interactive discussion, encouraging participants to share their experiences, perspectives, and insights. The session was guided by a skilled moderator (fourth co-author) who ensured that the conversation remained focused on key topics while allowing for flexibility to explore emerging themes. The discussions were recorded and also the main discussions were transcribed by the first two co-authors.

Participants included practitioners with diverse roles, ranging from project leads to senior ML scientists and data scientists to a senior digital adoption manager, providing a holistic view of the project. The focus group was structured around a set of predefined questions yet allowed for spontaneous discussions and follow-up questions to delve deeper into specific issues. This method is supported by established qualitative research practices, enabling the exploration of complex topics in a dynamic group setting.

During the session, participants discussed various aspects of the development process, including technical challenges, integration of nondeterministic LLMs, and strategies for ongoing validation and adaptive testing. Special attention was given to the principles of Responsible AI (RAI), with discussions on handling issues such as hallucinations, bias, and contextual precision.

By engaging in collective dialogue, the focus group uncovered critical insights and shared experiences that might not have emerged through other forms of data collection, e.g., individual interviews. This comprehensive approach allowed us to systematically map the software engineering challenges and procedures involved in constructing a RAG-based system, offering valuable guidance for both researchers and practitioners in the field. We conducted two brief follow-up sessions to address any uncertainties regarding the information discussed in the focus group and to redact information proprietary to Transurban.

## 5. Challenges and Future Research Directions

In this section, we present our study results sequentially, beginning with the challenges observed in the data ingestion workflow, proceeding to those identified in the response generation workflow, and concluding with challenges specific to the RAG evaluation process. For each challenge, we review the existing related literature and present research questions (RQs) that should be addressed by future research. Specific challenges associated with each step are summarized in Figure 2.

> **Challenge 1**–Multi-Modal Data Engineering for RAG-based Systems.

Data engineering plays a vital role as part of the software design phase for RAG-based systems (e.g., a data diagram). One of the primary challenges of building a RAG-based system at Transurban lies in handling multi-modal data within the data ingestion workflow. At Transurban, there are various types of data for effective tolling management systems. For instance, transaction data (e.g., toll charges, payment information), image/video data (e.g,. license plate information, vehicle types), user data (e.g,. account information, travel history), operational data (e.g., traffic flow, maintenance records), legal and policy documents, etc. These multi-modal data require embeddings that unify disparate data types, such as text processed by models like BERT and images by ResNet or CLIP. Such large-scale multi-modal data pose challenges to various research problems, e.g., what are the optimal approaches for data pipelines, data extraction, data indexing, information retrieval, and query performance.

**Existing Literature.** To handle multimodal data for RAG, prior studies [64, 65] proposed a multimodal transformer for learning cross-modal representation with an external memory to obtain multimodal knowledge contained in images or text snippets in order to augment the generation. Retrieval algorithms also play a vital role in managing large-scale multi-modal datasets by efficiently retrieving data for Retrieval-Augmented Generation (RAG). These algorithms vary in their approach and effectiveness depending on the specific requirements of the task at hand. For instance, Cosine similarity, TF-IDF, and BM25 [66] are among the most popular information retrieval algorithms that scores documents based on the presence and frequency of query terms, as well as the overall length of the documents. This method has proven to be highly effective for text retrieval tasks, ensuring relevant documents are prioritized in search results.

**Future Research Directions.** Although existing work focuses on multimodal transformer and retrieval algorithms, they are not designed for RAG systems and LLMs. Specifically, future research should address the research question as shown in 3.

Table 3: Summary of Research Challenges and Questions for RAG Systems

| Challenge | Research Questions |
|---|---|
| **Challenge 1– Multi-Modal Data Engineering for RAG-based Systems.** | RQ1.1 - How can we develop RAG systems that can handle large-scale multi-modal data, while achieving optimal query performance? |
| | RQ1.2 - How can we optimally extract key information from large-scale multi-modal data? |
| | RQ1.3 - How can we build multimodal knowledge index? |
| | RQ1.4 - How can we optimally create multimodal vector representation? |
| | RQ1.5 - How can we automatically test and measure the data quality for RAG systems? |
| **Challenge 2– Adaptive Security Guardrails to Protect Sensitive Input and Prevent Hacking.** | RQ2.1 - How can we develop adaptive guardrail frameworks that dynamically learn and adjust to new security threats in RAG systems? |
| | RQ2.2 - How can we effectively detect and adaptively respond to sensitive user inputs in RAG systems? |
| **Challenge 4– Balancing the Relevancy and Conciseness of the RAG-Generated Responses.** | RQ4.1 - How can we efficiently identify the optimal combination of hyperparameters to balance relevancy and conciseness in RAG-generated responses? |
| | RQ4.2 - How can alternative techniques beyond hyperparameter tuning contribute to balancing relevancy and conciseness in RAG-generated responses? |
| | RQ4.3 - What are the other quality aspects of the generated responses beyond the relevancy and conciseness? |
| **Challenge 5– Automated Testing for RAG-based Systems.** | RQ5.1 - How can we automatically generate test inputs and test oracle (including a benchmark dataset) for RAG-based systems at scale? |
| | RQ5.2 - How can we determine the quality, coverage, and adequacy of generated test cases for RAG-based systems? |
| | RQ5.3 - What are the novel characteristics of software defects for RAG-based systems? |
| | RQ5.4 - How can we automatically debug and repair RAG-based systems? |
| **Challenge 7– Analysing and Incorporating Human Feedback for Automated Continuous Improvement at Scale.** | RQ7.1 - What is the key usage interaction data that should be collected for developing usage analytics? |
| | RQ7.2 - What are the novel characteristics and taxonomy of human feedback for RAG-based systems? |
| | RQ7.3 - How can we automatically analyze, categorize, and summarize human feedback for RAG-based systems? |
| | RQ7.4 - How can we automatically extract key information from millions of human feedback and transform it into actionable insights or new features? |
| | RQ7.5 - Once actionable insights and new features are extracted, how can we automatically incorporate them into the RAG-based systems for automated continuous improvement? |
| **Challenge 8– Responsible AI for RAG-based Systems.** | RQ8.1 - What are the specific Responsible AI frameworks for RAG-based systems? |
| | RQ8.2 - How can we ensure that the generated responses from RAG-based systems adhere to Responsible AI principles? |
| | RQ8.3 - How can we develop RAI test automation tools that can evaluate the retrieval and generation components of RAG-based systems? |

**Challenge 2**–Adaptive Security Guardrails to Protect Sensitive Input and Prevent Hacking.

The use of a RAG system introduces privacy challenges when customers inadvertently input sensitive information, such as credit card or ID numbers, which undergo processing across input collection, prompt augmentation, and LLM response generation stages. This process exposes the system to risks like unauthorized data access or leaks during storage, interception in transit if unencrypted, and vulnerabilities in temporary memory storage. Beyond these concerns, RAG systems face vulnerability to targeted attacks like adversarial information retrieval, where attackers manipulate retrieval mechanisms to bias information generation. For instance, injecting false toll fee details into Transurban's customer virtual assistant could mislead users with incorrect payment instructions or inaccurate toll rates, compromising service reliability. Thus, these security concerns pose challenges to research problems, e.g., how to develop adaptive guardrail rules to mitigate both unintentional and intentional threats.

**Existing Literature.** Prior works have proposed security guardrails to mitigate security exposure such as prompt injection attacks and sensitive data breaches for mitigating these risks [67, 43]. Specifically, Rai et al. [67] introduced GUARDIAN (Guardrails for Upholding Ethics in Language Models), a multi-tiered defense architecture designed to protect LLMs from adversarial prompt attacks. The approach includes several key components: a system prompt filter, a pre-processing filter utilizing a toxic classifier and ethical prompt generator, and a pre-display filter using the model itself for output screening. Furthermore, Rebedea et al. [43] proposed NeMo Guardrails, an open-source toolkit for adding programmable guardrails to LLM-based conversational systems. This approach allows developers to add user-defined guardrails independently of the underlying LLM, ensuring flexibility and interoperability.

**Future Research Directions.** Although existing guardrail solutions have been proposed, they typically rely on rule-based systems with pre-defined guardrails. While effective in certain contexts, these systems often lack adaptiveness and flexibility to handle the diverse and evolving scenarios encountered in real-world applications. For instance, they may struggle with dynamically adjusting to new types of threats or understanding nuanced contextual changes in conversational settings. Specifically, future research should address the research question as identified in Table 3.

> **Challenge 3**–Managing and Operating the Latest LLMs Versions for RAG-Based Systems Without Sacrificing Its Performance.

Due to the fast-evolving landscape of LLMs, migrating the existing RAG-based virtual assistant to the latest LLM version introduces significant challenges. While newer LLMs include more up-to-date knowledge and enhanced generative capabilities, migrating to the latest LLM version involves ensuring that the new LLM works seamlessly with our existing retrieval mechanisms, preserving the accuracy and relevance of the retrieved data. For example, changes in how the new LLM interprets context or generates responses could affect how retrieved documents are utilized, potentially leading to inconsistencies or reduced effectiveness. Thus, the LLM transition poses specific research challenges, such as how to update the LLM component within RAG systems while ensuring backward compatibility with our retrieval system and maintaining the integrity of retrieved information.

**Existing Literature.** Mousavi et al. [68] developed a novel approach to dynamically evaluate the knowledge in LLMs and their time-sensitiveness. When tested on 24 open-source LLMs, the results imply that a significant portion of the models' outputs are either outdated or irrelevant, potentially leading to misinformation if relied upon.

**Future Research Directions.** To develop LLM applications that can handle the deployment of new versions of LLMs frequently, a robust LLMOps pipeline must be well-defined and set up properly to ensure changes in the LLM model do not affect other components in the pipeline. This includes components in the data ingestion pipeline, model fine-tuning and prompt engineering, and deployment and testing. By establishing such robust pipelines, Transurban may be able to mitigate the challenges of rapid model changes in the application, resulting in less effort in the maintenance of the virtual customer assistant.

> **Challenge 4**–Balancing the Relevancy and Conciseness of the RAG-Generated Responses.

Another notable challenge faced by Transurban is to balance the relevancy and conciseness of the responses generated by LLMs. For example, when the testing team sets a low temperature, the model generates responses such as: "Pay toll online at link." This response, while relevant, lacks sufficient detail and clarity for the customer. Conversely, when the testing team sets a high temperature, the model generates excessively verbose responses, nearly generating the entire customer support document. This response, while detailed, is too long and may overwhelm the customer.

**Existing Literature.** Peeperkorn et al. [69] used a narrative generation task to analyze LLM output across different temperature values. They evaluated LLM's outputs based on novelty, typicality, cohesion, and coherence. They found a weak correlation between temperature and novelty, a moderate correlation with incoherence, and no relationship with cohesion or typicality. Furthermore, Renze and Guven [70] investigated the effect of temperature on LLM performance in problem-solving tasks. They found that temperature adjustments within a certain range (0.0 to 1.0) do not significantly impact performance, suggesting that fine-tuning within this range may not resolve the issue.

**Future Research Directions.** Typically, practitioners often adjust the temperature aiming to achieve a balanced output. However, existing works have found that temperature adjustments alone have a limited impact. The findings also imply that other parameters (e.g., top-k sampling, top-p sampling, and repetition penalty [71]) or techniques might need to be explored to achieve the desired balance of relevance and conciseness. Thus, finding the optimal settings, which are not limited to tuning hyperparameters, to ensure responses are both relevant and concise poses a challenge to research problems, as shown in Table 3.

> **Challenge 5**–Automated Testing for RAG-based Systems.

Testing Retrieval-Augmented Generation (RAG) systems pose significant challenges primarily due to the lack of a ground truth oracle (i.e., how do we know if the RAG-generated responses are correct). At Transurban, the engineering team expended substantial effort merely to develop basic test cases to ensure the fundamental functionalities of their RAG-based customer virtual assistant. However, the large language models behind RAGVA often struggled to interpret and respond accurately to various customer utterances, sometimes even replying in unexpected language (e.g., ask in English but reply in French, prompt "chat to a human" but the conversation may be incorrect transferred to an LLM agent instead of a human agent). In the absence of an established ground truth, evaluating the correctness of these responses becomes challenging.

**Existing Literature.** Various automated testing approaches [72] have been proposed for traditional software systems and ML-based software systems. For example, test input generation [73, 74, 75, 76], test oracle generation [77], test case prioritization [78, 79, 80], test case reduction [81], test case quality and adequacy [82, 83], test coverage measures [84], debugging and repairs [85], bug management [86, 87], etc. However, they fall short when it comes to testing RAG-based systems.

**Future Research Directions.** Future research should address the RQs described in Table 3.

> **Challenge 6**–Establishing Comprehensive and Systematic Evaluation Metrics for RAG-based Systems.

To evaluate the performance of the RAG virtual customer assistant, Transurban has been experimenting with a variety of metrics commonly used on AI models. Traditional NLP metrics such as BLEU [44] or ROUGE [45] may not provide the best representation since they can often score quite low for sentences that have the same meaning [88]. Comprehensive metrics for RAG systems that effectively measure both retrieval and generative performance are currently lacking and need further research.

**Existing Literature.** Zhang et al. [89] introduced BERTScore, an evaluation metric that leverages pre-trained BERT contextual embeddings to measure the similarity between two sentences by summing the cosine similarities of their token embeddings. This approach offers a nuanced assessment of semantic similarity, making it particularly suitable for RAG models. In evaluations involving machine translation and image captioning tasks, BERTScore demonstrated stronger correlations with human judgments at both the system and segment levels. Liu et al. [90] developed G-EVAL, a framework utilizing large language models with chain-of-thought (CoT) reasoning and a form-filling paradigm to assess the quality of natural language generation systems. G-EVAL generates CoT evaluation steps based on the task and evaluation criteria, then uses these to evaluate the LLM's output for a given input text. When tested on the Topical-Chat benchmark [91], G-EVAL significantly outperformed previous state-of-the-art evaluators, aligning more closely with human judgments on aspects such as naturalness, coherence, engagingness, and groundedness.

**Future Research Directions.** Traditional language model evaluation metrics like BLEU, ROUGE and Cosine Similarity, often lacks the ability to evaluate the distinct characteristics of RAG models, such as Noise Robustness, Negative Rejection, and Counterfactual Robustness [23]. While more advance metrics such BERTScore and G-EVAL can provide more contextualised evaluation, they often involve using language models directly or through model embeddings, creating additional layers in the evaluation step that can be expensive in computational resources. Thus, future research should focus on investigating what are the comprehensive and systematic evaluation metrics for evaluating the performance of RAG-based systems.

> **Challenge 7**– Analysing and Incorporating Human Feedback for Automated Continuous Improvement at Scale

Like traditional software systems, RAGVA requires ongoing maintenance and continuous improvement through human feedback or user interaction data. To enable users to provide feedback on the RAGVA, a practical approach is to offer thumbs up/thumbs down buttons to indicate their satisfaction, along with the option for users to provide detailed feedback through comments. These comments can elaborate on their experiences, offer suggestions, or describe issues encountered when interacting with the RAGVA. However, given the large volume of feedback expected from users, the challenge lies in categorizing and summarizing human feedback, extracting key information, creating taxonomies of usability issues, and ultimately generating actionable insights or new features for continuous system improvement.

**Existing Literature.** Prior studies proposed various automated feedback analysis approaches for mobile apps, traditional software systems, and etc. For example, Tushev et al. [92] proposed an unsupervised approach for extracting useful information from mobile app reviews. Hadi and Fard [93] utilized pre-trained LLMs to classify app reviews to identify potential issues. Asnawi et al. [94] combined the contextualized topic model (CTM) and pre-trained language model to create a context-aware topic model that enhances the understanding of user experiences and opinions. Panthum and Senivongse [95] studied the methodology to utilize human feedback in creating functional requirements for mobile apps. Yan et al. [96] developed a framework to enhance the robustness of RAG-based systems by evaluating the quality of retrieved documents with an additional annotator, significantly improving the ability of automatic self-correction and efficient utilization of retrieved documents. Chen et al. [97] proposed a novel continual learning-based framework to incrementally incorporate new documents over time in the RAG pipeline, without having to retrain from scratch or losing the ability to retrieve previously learned documents.

**Future Research Directions.** Due to the different nature of the human feedback for RAGVA (i.e., why generated responses are incorrect) and mobile apps and traditional software systems (i.e., why a mobile app is not useful), existing automated feedback analysis approaches may not be applicable. Automating RAG-based systems to continuously learn from human feedback remains in its early stages of development, thus, future research should address the RQs outlined in Table 3.

> **Challenge 8**–Responsible AI for RAG-based Systems.

Transurban has implemented a rule-based guardrail system to block malicious or inappropriate prompts from reaching the LLM, effectively preventing harmful content generation. However, existing off-the-shelf tools from cloud providers (AWS, Azure, Google) only provided limited options to evaluate whether the RAG-generated responses that adhere to RAI principles, such as preventing hallucination, ensuring answer relevancy, detecting toxicity and mitigating bias. This presents a challenge for Transurban to ensure the responsible use of virtual assistant.

**Existing Literature.** As presented in Table 1, numerous RAI metrics have been defined and investigated by previous literature, such as Hallucinations [49], Faithfulness [50], Bias [51], Toxicity [52], Contextual Precision/Recall [53, 54], Contextual Relevancy [55] and Knowledge Retention [56]. Automated tools such as DeepEvalAI [60] also provides evaluation techniques to rigorously evaluate LLM outputs on the above-mentioned metrics. Utilizing these automated tools are crucial in preventing the generation of harmful or inappropriate content from the RAGVA at Transurban.

**Future Research Directions.** Although current automated tools offer certain RAI evaluation for LLMs, RAG-specific RAI frameworks have received little attention. For example, when using a framework to detect hallucinations in LLM responses, the same approach might be excessive for evaluating hallucinations in RAG systems. Unlike plain LLMs, RAG systems base their generated responses on a knowledge base, which means the methodology for detecting hallucinations in RAG systems should consider this existing reference material, making the process inherently different. To address the lack of RAI frameworks for RAG, future research should address the RQs presented in Table 3.

## 6. Practical Implications and Consequences of Unaddressed Challenges

Understanding the potential consequences of leaving these challenges unresolved is essential, as it underscores the risks to the effectiveness, scalability, and trustworthiness of RAGVAs. Below, we outline these practical implications for software practitioners.

- Challenge 1 relates to the limitations in use case and document requirements of the system, where software practitioners must consider the scope of the RAG system. Without robust solutions for multi-modal data engineering, RAGVAs may struggle to process diverse data types efficiently, leading to limited scalability towards future use cases and reduced query performance from incomplete retrieval of critical information [98].

- Challenge 2 addresses the potential security aspects of the system, such as preventing adversarial attacks or unauthorized access to sensitive information. Without robust security mechanisms, RAGVAs may expose organizations to data breaches, legal liabilities, and reputational damage. For example, studies on adversarial retrieval highlight how injected misinformation can lead to inaccurate responses, compromising the trustworthiness of the system [99].

- Challenge 3 addresses the issue of post-deployment maintenance of RAGVAs; as LLMs are a crucial component in the framework, frequent updates and rollouts mean engineers have to fine-tune the model for every new iteration – often time-consuming without a robust LLMOps pipeline.

- Challenge 4 relates to the fine-grain adjustments of the verbosity of the system, as practitioners aim to find the balance between providing relevant information and concise outputs to satisfy end-users. Without fine-tuning LLM parameters effectively, practitioners risk producing verbose, uninformative, or incomplete responses, which frustrates users and undermines the system's utility [100].

- Challenge 5, the implementation of automated testing can validate RAGVA after every iteration of development, and the lack of such a framework would mean much manual work and repeated testing procedures performed by developers due to the added complexity of additional components compared to traditional LLM testing [101].

- Challenge 6 emphasizes the need for more comprehensive metrics to ensure the quality and relevance of the RAGVA outputs with less commitment of time and resources. Testing conducted on limited metrics would lead to shallow system evaluation, as effective testing requires a holistic approach that addresses all components within the RAGVA, ranging from document ingestion to response generation [38].

- Challenge 7 is associated with post-deployment and continuous system improvement, as practitioners aim to use human feedback to adjust the RAGVA. If feedback mechanisms are underdeveloped, practitioners may struggle to extract actionable insights, leaving little room for continuous improvement [102].

- Finally, challenge 8 is vital for ensuring the ethical and legal use of LLMs in RAGVA. Practitioners are required to follow laws and requirements in their respective regions, for example the EU AI Act [103], or data privacy acts such as the GDPR [40]. Hence, RAI principles are crucial in aligning RAGVAs with such ethical and legal requirements, where failing to implement these damages trust and lead to societal harm [25].

## 7. Discussions

In this section, we discuss some fine-grain details of developing and maintaining the RAGVA. This includes guidance on the choices made within Section 3, as well as alternatives that may benefit practitioners in developing their own RAG-based systems.

### 7.1. Chunking Strategy

Document chunking strategy can affect the performance of RAG systems, and should be considered carefully based on several criteria such as document type, content density, and the intended use case of the system. For single-modal text documents, fixed-size chunking (e.g., 100, 256, or 512 tokens) is a common approach, offering simplicity and consistency but risking truncation of sentences or inclusion of excessive noise. Recursive and sliding-window chunking strategies address these issues by dynamically adjusting chunk boundaries based on linguistic cues like punctuation, preserving semantic coherence. Semantic or agentic chunking, which involve NLP techniques or LLMs can dynamically structure chunks, ideally for highly nuanced and unstructured documents but may be computationally expensive. For specialized documents like reports, element-based chunking, which preserves logical structures such as tables and headings, ensures the retention of meaningful content and hierarchy. In scenarios requiring time-sensitive retrieval, metadata-enriched chunks with timestamp or summaries enable efficient, context-aware filtering. Selecting the optimal chunking strategy depends on document complexity, structure, and the contextual depth required to answer specific queries effectively [23, 104].

To incorporate multi-modal data into the RAGVA, such as visual, audio, and video documents, or beyond, these documents can either be converted into text data or processed using multi-modal retrieval frameworks. For text conversion, approaches such as Optical Character Recognition (OCR) models, namely Google's Tesseract [105], which can extract text from images. Automatic Speech Recognition (ASR) models provide highly accurate transcriptions, while video-to-text models offer contextual information about scenes in addition to transcribing video content. Furthermore, multi-modal RAG pipelines have also been developed. For instance, the RA-CM3 model [98] retrieves both text and images from the knowledge base to generate responses. This capability could enhance the RAGVA by enabling it to produce customer form examples or visual guidelines for navigating the user portal.

*7.2. Embedding Model and Vector Database Selection*

The selection of embedding models and vector databases in the ingestion & generation phase of the RAGVA can significantly affect the system's retrieval accuracy [106].

In terms of the embedding model, their semantic representation capability plays a key role in solving the semantic space-matching problem between queries and chunk blocks. Static embedding models such as Word2Vec, GloVe, TF-IDF, and BM25 are usually utilized for sparse retrieval methods, which are effective for exact or lexical matches but struggle with capturing semantic similarity [101]. For dense retrieval methods, which are more suited for tasks requiring deeper semantic understanding, pre-training language models (PLMs) such as BERT, OpenAI Embeddings or T5-based models can be used to handle diverse user queries and complex document structures [23]. These embedding models also enable additional fine-tuning to be conducted to optimize performance for particular applications.

In developing RAG-based systems, selecting an embedding model can be challenging as several factors need to be considered such as retrieval accuracy and scalability. Previous works have conducted performance benchmarks that include metrics such as Mean Reciprocal Rank (MRR), Precision@K, and Recall@K, which enable comparisons of retrieval accuracy across domain-specific datasets [107, 108]. Additionally, the scalability of the model to handle large-scale data efficiently and with low inference latency is critical, especially in high-query-volume systems. Lightweight embedding models such as DistilBERT or MiniLM may better suit responsiveness demands of real-time applications without compromising retrieval accuracy. Furthermore, it is equally important for the embedding model to be able to capture the sematic nuances of the target domain, where task-specific fine-tuning or the use of task-specific models can be effective at achieving optimal performance [65].

When selecting vector databases to store the embedded vectors, factors such as flexibility, efficiency and scalability need to be considered to facilitate various indexing and nearest neighbors methods. Key selection criteria include support for multiple index types, allowing optimization for diverse data characteristics and use cases; billion-scale vector support, critical for handling large datasets in LLM applications; hybrid search capabilities, which combine vector search with traditional keyword search to enhance retrieval accuracy; and cloud-native features, ensuring seamless integration, scalability, and management in cloud environments [106]. To evaluate selected vector databases, practitioners may use metrics including cosine similarity, Euclidean distance, each offering unique insights into the effectiveness of similarity matching [109].

*7.3. Test Case Generation and Maintenance*

Besides obtaining test cases manually, additional test cases could be sourced from various open-source RAG benchmark datasets, as highlighted in Table 1. For example, QA datasets such as Natural Questions [47] and WikiEval [110] provide both the correct answer to each question, as well as the relevant context passages, to evaluate the retrieval and generative capabilities of RAGVA. Furthermore, LLMs can also be employed to automatically generate diverse and contextually relevant test cases based on existing ones, where they can be instructed to prepare questions and answers of similar nature, given the relevant documents. We provide example test cases for evaluating these aspects in Table 2.

To maintain the test cases during automated testing, evolving system requirements and domain-specific use cases need to be analyzed. Alégroth et al. [111] evaluated the cost of maintaining software tests and found that developing new tests is costlier than maintenance, but frequent maintenance is less costly than infrequent, big bang maintenance. For RAG-based systems, practitioners should prioritize continuous test case updates to align with the intended use cases and system objectives. Additionally, if the data sources retrieved by the RAGVA are modified, such as documentation updates or structure changes, test cases must be updated accordingly. This ensures that system performance evaluations remain accurate and relevant, highlighting the critical role of automated testing pipelines in efficiently managing such updates and minimizing the risks of system drift or outdated evaluations.

## 8. Threats to Validity

**Threats to construct validity** relate to the assumptions underlying the study's focus on RAG and LLMs for VAs. While the identified challenges and solutions are grounded in real industrial settings, they may not fully generalise across different domains or use cases. In particular, assumptions about the applicability of RAGVA to diverse VA scenarios may introduce bias if the underlying requirements differ significantly from the studied use case at Transurban. To mitigate this, the challenges we derived in this study are not specific to Transurban's context thus the findings and future research directions are still applicable to a broad range of RAG-based applications.

**Threats to internal validity** relate to the reliance on focus groups as the primary method of data collection. While focus groups can generate broad insights, they may lack the depth of individual interviews where more detailed and personal information can be explored. Further, the setting and context of the focus group can influence participants' behavior and responses. To mitigate these issues, we first had an experienced moderator, who's an expert in requirements elicitation using focus groups and has done several such sessions in the past. We also provided the option of follow-up chats to all nine participants.

**Threats to external validity** relate to the generalizability of our findings. Our study is rooted in Transurban's development of RAGVA, which may limit the applicability of the findings to other domains, industries, or project scales. The challenges identified in the paper reflect the priorities and circumstances specific to Transurban and our nine practitioner participants. While the identified challenges in Section 5 are supported by existing literature, they may not comprehensively represent all challenges in developing RAGVA or similar systems. We further note the generalizability of our work concerning the rapidly evolving nature of RAG and LLM technologies for VAs. As these technologies advance, the limitations and challenges identified in our research may change, potentially impacting the applicability of our findings and proposed future directions. To mitigate this, we have based our findings on real industrial settings and abstracted them to ensure they remain relevant despite potential updates in LLM technologies. More studies in broader and diverse contexts are required to improve the external validity of the study.

## 9. Conclusion

In this paper, working with Transurban, we present a comprehensive framework for building and evaluating a RAG-based virtual assistant. This framework is crucial for researchers and practitioners as it provides a structured approach to navigating the complexities of developing such advanced systems. Furthermore, we conducted a focus group study with nine members of the Transurban engineering team to gain in-depth insights into their development processes and identify specific challenges they encountered. Finally, we identified eight key software engineering challenges in building and evaluating a RAG-based virtual assistant and recommend future key research directions that are relevant to real-world contexts.

**Acknowledgement**

We thank Transurban and the CSIRO Next Generation Graduate AI Program: Creating Responsible AI Software Engineering Capability (RAISE) for their support and collaboration. Additionally, we appreciate the valuable contributions of all focus group participants. The perspectives and conclusions presented in this study are solely the authors' and should not be interpreted as representing the official policies or endorsements of Transurban or any of its subsidiaries and affiliates. Additionally, the outcomes of this study are independent of, and should not be construed as an assessment of, the quality of products offered by Transurban.